# Equivalent Circuit Analysis of Super-Planckian Emission into Far Field

Stanislav I. Maslovski[1], Constantin R. Simovski[2]
[1]Instituto de Telecomunicações, Universidade de Coimbra, Coimbra, Portugal
[2]Department of Radio Science and Engineering, Aalto University, Espoo, Finland
stas@co.it.pt

## ABSTRACT

Using our recently developed equivalent circuit model of radiative heat transfer we analyze the far-field thermal emission from bodies of constrained dimensions. We prove that the power radiated by a hypothetical metamaterial emitter which is conjugate matched with all the harmonics of the emitted field at a given wavelength can be infinitely higher than the power emitted by an ideal black body of the same radius. However, for optically large bodies, fulfilling the conjugate match conditions for higher-order harmonics is not a trivial task, which effectively limits the far-field radiated power by the well-known Planck law.

## 1. Introduction

Recently, there appeared works that deal with super-Planckian emission in the far-field. Of a particular interest is the question whether such emission can be obtained in the case of an optically large body with characteristic dimensions much greater than the radiation wavelength $\lambda$. For optically small bodies, Rytov, in his seminal work on fluctuational electrodynamics [1], has shown that the thermal emission power radiated by an emitter with characteristic radius $a \ll \lambda$ can be much greater than the power emitted by an ideal black body of the same radius. Such a super-Planckian emitter can be physically realized as a short antenna tuned at resonance with a conjugate matched load: $Z_{\rm load} = Z_{\rm in}^*$, where $Z_{\rm in}$ is the antenna input impedance. The increase in the radiated power is explained in this case by the fact that a small resonant antenna may have an effective area (i.e., the effective scattering cross-section) much greater than the square of the characteristic diameter of the antenna.

The thermal emission from an optically large body with characteristic radius $a \gg \lambda$ can be enhanced by placing it in a transparent shell with radius $r > a$ and with refractive index $n > 1$. In this case the increase in the emitted thermal radiation power can be explained by optical magnification effect: The size of the emitter as is seen through the transparent shell is larger (about $n$ times larger when $r \gg a$), which increases emitter's effective area. However, the physical size of such an enhanced radiator is also greater than the original size of the emitter without the shell. It is clear that the emitted power in this structure may not exceed the power radiated by an ideal black body with the radius equal to the outer radius of the shell.

In contrast, in this work we look for a possibility to obtain super-Planckian radiation from an optically large emitter with a constrained size. We fix the radius of the emitter and optimize its electromagnetic properties so that the emitted power is maximized. Then, we compare the thermal emission of this electromagnetically optimal emitter with the emission of the ideal black body with the same radius. It is important to note here that we do not demand the surface reflectance of such an optimal emitter to vanish for arbitrarily incident plane waves. Because of this, our optimal solution is conceptually different from the standard Kirchhoff's black body case.

## 2. Equivalent Circuit of Far-Field Thermal Emission

Recently, we have developed an equivalent circuit approach [2] that reduces full-wave thermal emission problems to circuit theory problems operating with effective fluctuating voltages and currents instead of the electromagnetic fields. This is achieved by expanding the radiated field at a given frequency into a set of suitable spatial harmonics. In this approach, the electromagnetic interaction of a hot emitter with the surrounding space is modeled at each spatial harmonic by the equivalent circuit shown in Fig. 1.

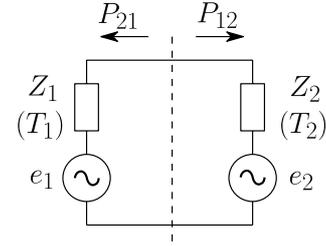

Fig. 1: The equivalent circuit of radiative heat transfer between an emitter equivalently represented by the complex impedance $Z_1(\omega)$ and its environment represented by the complex impedance $Z_2(\omega)$.

In this circuit, $Z_1(\omega)$ represents the equivalent complex impedance of emitter's body for a given harmonic. Respectively, $Z_2(\omega)$ is the equivalent complex impedance of the surrounding space for the same harmonic, which, in case of the free space, is simply the wave impedance of the corresponding mode: $Z_2 \equiv Z_{\rm w}$. We use time dependence of the form $\exp(j\omega t)$, with $j = \sqrt{-1}$. The thermal fluctuations in this circuit are taken into account by a pair of fluctuating electromotive forces $e_1(\omega)$ and $e_2(\omega)$, with mean square magnitudes (here, $i = 1, 2$)

$$\frac{d\overline{e_i^2}}{d\omega} = \frac{2}{\pi} \frac{\hbar\omega \, {\rm Re}(Z_i)}{e^{\frac{\hbar\omega}{k_{\rm B}T_i}} - 1} \qquad (1)$$

This result follows from application of the fluctuation-dissipation theorem to the circuit of Fig. 1,

and coincides with Nyquist's formula for the thermal noise in electric circuits. Here, $T_{1,2}$ are the temperatures of the emitter and the surrounding space, respectively.

The power delivered within a narrow range of frequencies $\omega \pm d\omega/2$ from the side of the emitter, $Z_1$, to the side of the environment, $Z_2$, is expressed in this formulation (per each spatial harmonic) as

$$dP_{12} = \frac{2}{\pi} \frac{\hbar \omega \, d\omega}{e^{\frac{\hbar \omega}{k_B T_1}} - 1} \frac{\text{Re}(Z_1)\text{Re}(Z_2)}{|Z_1 + Z_2|^2} \qquad (2)$$

As is clearly seen from Eq. (2), for the modes with a non-vanishing real part of the wave impedance: $\text{Re}(Z_2) > 0$, the delivered power is maximized under the conjugate match condition: $Z_1 = Z_2^*$. Under this condition, the absolute maximum of the emitted power per a spatial harmonic per unit of frequency is

$$\frac{dP_{\max}}{d\omega} = \frac{1}{2\pi} \frac{\hbar \omega}{e^{\frac{\hbar \omega}{k_B T_1}} - 1} \qquad (3)$$

Note that this expression is independent of $Z_{1,2}$. Due to orthogonality of the spatial harmonics, if the radiated power is maximized separately at all such harmonics, the total emission is also maximized.

## 3. Far-Field Super-Planckian Emission

We consider an arbitrary thermal emitter that is fully contained within a sphere of fixed radius $r = a$. The radiated field of this emitter can be expanded over the complete set of vectorial spherical waves, defined with respect to a spherical coordinate system $(r, \theta, \varphi)$ whose origin lies within this volume. These modes split into TE waves (with $E_r = 0$) and TM waves (with $H_r = 0$), with the field vectors expressed through a pair of scalar potentials $U_{nm}, V_{nm} \propto R_n(k_0 r) Y_n^m(\vartheta, \varphi)$, where $Y_n^m(\vartheta, \varphi)$ are Laplace's spherical harmonics, and $R_n(x) = \sqrt{\pi x / 2} \, H_{n+1/2}^{(2)}(x)$. The transverse electric field $\mathbf{E}_t = (0, E_\vartheta, E_\varphi)$ and the transverse magnetic field $\mathbf{H}_t = (0, H_\vartheta, H_\varphi)$ in these modes are related as $\mathbf{E}_{t,nm} = -Z_{w,nm}^{TE,TM} (\hat{\mathbf{r}} \times \mathbf{H}_{t,nm})$ where $Z_{w,nm}^{TE,TM}$ is the wave impedance of the spherical wave harmonic with the polar index $n$ ($n = 1, 2, \dots$) and the azimuthal index $m$ ($m = 0, \pm 1, \pm 2, \dots, \pm n$), which can be expressed as

$$Z_{w,nm}^{TE} = \eta_0 \frac{j R_n'(k_0 r)}{R_n(k_0 r)} \qquad (4)$$

$$Z_{w,nm}^{TM} = \eta_0 \frac{j R_n(k_0 r)}{R_n'(k_0 r)} \qquad (5)$$

Note that the wave impedance of a mode depends on the radial distance $r$ and the polar index $n$, and it is independent of the azimuthal index $m$. We exclude the purely longitudinal mode with $n = m = 0$ because it does not contribute to the radiated power.

Note that the spherical wave impedance given by Eqs. (4,5) has non-vanishing real part for the harmonics with arbitrary high indices $n$ and $m$. Thus, there are no fully evanescent waves among the spherical wave harmonics: Each mode, whatever high index it has, contributes into the far field. From here one may see that at any given wavelength there is a possibility, at least a purely theoretical one, to satisfy conjugate match condition for the entire spectrum of spherical waves that are emitted by a body with a finite radius. Under this, the total power (per unit of frequency) emitted by the body at this wavelength, is [using Eq. (3)]

$$\frac{dP_{\text{tot}}}{d\omega} = 2 \times \sum_{n=1}^{\infty} \sum_{m=-n}^{n} \frac{1}{2\pi} \frac{\hbar \omega}{e^{\frac{\hbar \omega}{k_B T_1}} - 1} \to \infty \qquad (6)$$

i.e., it grows infinitely. Thus, at least from a purely theoretical point of view, there is no upper limit on the far-field thermal emission from a spherical body of a fixed radius.

If we exclude the strongly reactive modes with indices $n > N_{\max} = 2\pi a / \lambda$, $N_{\max} \gg 1$, from the summation (6), we obtain for the spectral density of power

$$\frac{dP_{\text{tot}}}{d\omega} = \frac{4\pi a^2}{\lambda^2} \frac{\hbar \omega}{e^{\frac{\hbar \omega}{k_B T_1}} - 1} \qquad (7)$$

Recognizing $4\pi a^2$ as the area of the spherical surface, we obtain from Eq. (8):

$$\frac{dP_{\text{tot}}}{d\omega \, dS} = \frac{\hbar \omega^3}{4\pi^2 c^2 \left( e^{\frac{\hbar \omega}{k_B T_1}} - 1 \right)} \qquad (8)$$

which coincides with the result for the spectral density of power emitted by unit surface of a black body. Thus, super-Planckian emission into the far field can be achieved if and only if the emitter efficiently excites higher-order spherical harmonics, which are not radiated from the black body emitter.

## 4. Concluding remarks

To conclude, let us note that realizing the conjugate match condition for waves with $n > N_{\max}$ in a practical thermal emitter is not a trivial task. First of all, it is obvious that it is impossible to achieve such matching in emitters formed by homogeneous dielectrics or magnetics. Perhaps, a practical possibility lies in using media (e.g., metamaterials) that are inhomogeneous on the scale of the characteristic spatial variation in the near-field of a given spherical harmonic. The difficulties that one would have to overcome in such designs should be of the same degree as when realizing superdirective antennas.